# Topological Lifshitz transition in Weyl semimetal NbP decorated with heavy elements

Ashutosh S Wadge,[1,*] Bogdan J Kowalski,[2,*] Carmine Autieri,[*,1] Przemysław Iwanowski,[1,2] Andrzej Hruban,[2] Natalia Olszowska,[3] Marcin Rosmus,[3] Jacek Kołodziej,[3] and Andrzej Wiśniewski[1,2]

[1]*International Research Centre MagTop, Institute of Physics, Polish Academy of Sciences, Aleja Lotników 32/46, PL-02668 Warsaw, Poland*
[2]*Institute of Physics, Polish Academy of Sciences, Aleja Lotników 32/46, PL-02668 Warsaw, Poland*
[3]*National Synchrotron Radiation Centre SOLARIS, Jagiellonian University, Czerwone Maki 98, PL-30392 Cracow, Poland*

Studies of the Fermi surface modification after in-situ covering NbP semimetal with heavy elements Pb and Nb ultrathin layers were performed by means of angle-resolved photoemission spectroscopy (ARPES). First, the electronic structure was investigated for pristine single crystals with two possible terminations (P and Nb) of the (0 0 1) surface. The nature of the electronic states of these two cleaving planes is different: the P-terminated surface shows spoon and bow tie-shaped surface states, whereas these shapes are not present in the Nb-terminated surface. ARPES studies show that even 1 monolayer (ML) of Pb causes topological Lifshitz transition (TLT) in P-terminated NbP where the surface Fermi arcs (SFAs) teleport to another pair of Weyl points connecting two adjacent Brillouin zones. Depositing 1 ML of Pb modifies the Fermi surface along with a shift in the Fermi energy. On the other hand, the deposition of approximately 0.8 ML of Nb modifies the electronic structure of P-terminated NbP, pushing the system on the verge of TLT but not yet fully transformed. Regardless of the dramatic surface evolution, SFAs remain connected to topologically protected Weyl points (WPs). Additionally, we studied the Nb-terminated NbP covered with 1.9 ML of Pb with only altered trivial surface states caused by an ordinary Lifshitz transition.

## I. INTRODUCTION

From the point of view of the electronic structure of topological Weyl semimetal (WSM), the most relevant feature is the existence of open surface Fermi arcs (SFAs), emerging at the surface [1, 2]. Their interplay with bulk Weyl nodes may influence the emergence of Majorana modes when WSM is coupled with a superconductor [3, 4]. These intriguing phenomena mainly involve SFAs, hence from the point of view of basic research and applications, it is very relevant to find an effective way of controlling and modification of SFAs e.g. of switching them between pairs of Weyl points (WPs). WSMs are possible only in systems without a center of inversion or when the time-reversal symmetry is broken. WSMs can be divided into two classes (type-I and type-II), depending on whether the system conserves or violates Lorentz symmetry. Angle-resolved photoemission spectroscopic measurements have shown that e.g. NbP and TaAs are type-I WSMs [5, 6] whereas e.g. WTe$_2$ is type-II WSM [7, 8]. Some topological semimetals, as TaAs$_2$, become type-II WSMs when an applied magnetic field breaks the time-reversal symmetry [9-11]. The type-I WSMs can be considered as direct negative band gap semiconductors. While the type-II ones exhibit an indirect negative gap [12]. In general, the Lifshitz transition is a change in the topology of the Fermi surface without breaking of any symmetry [13, 14]. Topological Lifshitz transition (TLT) is a consequence of external disturbances applied to topological materials due to which the Fermi surface feels the transition. The Lifshitz transition point is the critical point that divides the two phases of the system

For instance, bulk protected WPs, SFAs and their interconnections to WPs together determine the stability of the Fermi surface. In WSM, it occurs when e.g. Weyl cone is tilted so much that it tips over and crosses through the original Fermi level. The SFAs, which are also known as topologically non-trivial Fermi surface states (TOPOSS), are protected by bulk Weyl points. Hence, TOPOSS are robust against the weak perturbations applied to the system.

Recently, TLTs and the possibility of the Fermi arc manipulation were reported for NbAs Weyl semimetal decorated with light elements. Yang *et al*. demonstrated that by using in-situ surface decoration, always with light elements, it is possible to change the shape, size and connections of the Fermi arcs [15]. The topological Lifshitz transition moves the Fermi arcs; however, the Fermi arcs are still tied to a pair of Weyl points of opposite chirality, as is imposed by the bulk topology. While many topological Lifshitz transitions are obtained theoretically [16-18], their experimental observation is still rare. Recently, a published article with NbP Weyl semimetal thin films and chemical doping has shown the possibility of manipulation and control of the Fermi level [19].

Niobium mono-phosphide (NbP) is a noncentrosymmetric WSM containing 12 pairs of Weyl nodes in the Brillouin zone [20]. Freshly cleaved surfaces of NbP can be cationic or anionic in nature. The single crystals of NbP show trivial surface states as well as topologically non-trivial Fermi arcs. It is easy to change trivial surface states by covering the cleaved surface of NbP. Topologically protected Weyl fermions are highly mobile due to robust topological states which are stable against small perturbations whereas the chemical potential is very sensitive to small perturbations as in NbP.

[*]wadge@magtop.ifpan.edu.pl,
[*]carmine.autieri@magtop.ifpan.edu.pl

The two important reasons justify selecting Pb and Nb as the elements that covered the NbP surface. First, both Pb and Nb are superconductors at helium temperatures. They were used by us in the studies of conductance spectra of (Nb, Pb, In)/NbP - superconductor/Weyl semimetal junctions [21]. It turns out that Pb-NbP and Nb-NbP interface conductance values are much smaller in comparison with In-NbP, indicating strongly reduced active junction areas. However, their differential conductance spectra as a function of voltage show pronounced increases in the subgap regions. In these experiments, the deposited layers were several hundred nm thick. It was important to clarify what changes appear during the first stages of formation of those interfaces even in their normal state. On the other hand, as mentioned, other group reported the possibility of Fermi arc manipulation for NbP covered with light elements [19]. This gave the motivation to check the impact of heavy elements, which additionally are characterized by larger spin-orbit coupling and larger strength of the electronic hybridization (intended as the size of the hopping parameters).

In this paper, on the basis of angle-resolved photoemission spectroscopy (ARPES) studies, we report the influence of deposited heavy metals (Pb and Nb) on P- and Nb-terminated surfaces with observed topological Lifshitz transition appearing when Weyl semimetal NbP is covered with a 1 ML of Pb on P-terminated surface.

## II. EXPERIMENTAL DETAILS

### A. Single crystal growth of NbP and structural characterization

All details concerning single crystals growth and characterization were published in [21]. For the growth of NbP crystal, we applied chemical vapor transport (CVT) method. Crystal structure and crystallographic quality of grown NbP were verified by X-ray powder diffraction (XRD). It was found that NbP crystallizes in a body-centered tetragonal unit cell and its space group is I41md (No. 109). The lattice constants are: a = 3.33544(6) Å, c = 11.3782(3) Å and V = 126.584(4) Å$^3$. The energy-dispersive X-ray spectroscopy (EDX) confirmed that NbP crystals are well stoichiometric (within experimental error the atomic ratio was 1:1). The crystal structure of NbP with two possible cleavage planes of the crystal structures with Nb and P terminations are shown in Fig. 1 (a).

### B. Angle-resolved photoemission spectroscopy and deposition of metals (Pb, Nb)

The Pb and Nb deposition procedures as well as the ARPES measurements at each stage of deposition were carried out in the multi-chamber experimental setup of the ultra angle-resolved photoemission spectroscopy (UARPES) beamline at the National Synchrotron Radiation Centre SOLARIS in Krakow (Poland). The high energy and angular resolution of the spectrometer (1.8 meV and 0.1°, respectively) enabled the band mapping within the whole Brillouin zone with a precision sufficient for observation of the evolution of the band structure of NbP. The samples were kept at the temperature of 80 K. An elliptically polarizing quasiperiodic undulator of APPLE II-type was the source of the radiation for the experiments. The available photon energy range was 8–100 eV. The acquired ARPES data were analyzed with the use of the software procedure package developed by the UARPES beamline staff.

Sample surfaces suitable for ARPES experiments were obtained by cleaving the NbP monocrystals under UHV conditions. The crystallographic orientation of the cleaved samples was assessed in a separate XRD experiment. Then, after taking the reference ARPES spectra from the pristine surface, the first amount of metal (Pb or Nb) was deposited. The metal sources, a Joule heated evaporator for Pb and an electron beam evaporator for Nb, were installed in the preparation chamber interlocked with the electron spectrometer chamber. So, the evaporation processes left the vacuum in the spectrometer unimpaired. The deposition rate was estimated by preliminary tests with the quartz balance and independently assessed by core level measurements.

A custom-made Joule heated source evaporator was used to deposit Pb on in-situ freshly cleaved (0 0 1) surface of NbP. An evaporator is assembled in such a way that source Pb was held in a Tungsten coil. In order to attain the melting point of Pb, this coil was subjected to a DC current of a few Amps. The deposition process was done in an ultra high vacuum preparation chamber of the ARPES system.

Evaporator PREVAC EBV 40 A, which is specially made to grow monolayer thin films by molecular beam epitaxy process, was used for evaporation of Nb. A pure Niobium wire (Alfa Aesar, 0.5 mm, 99.06 %) was used as source material. In this system, an electron beam heats the Nb wire only at the end. The measured ion flux (proportional to the flux of evaporated atoms) is controlled by setting the electron emission current. Nb was deposited on in-situ cleaved (0 0 1) surface of NbP in ultra high vacuum.

## III. RESULT AND DISCUSSION

### A. Electronic structure of pristine NbP

The studied crystals were cleaved in-situ to obtain (0 0 1) surfaces with two possible terminations (P and Nb) in different trials of the cleaving processes. Surface-sensitive ARPES at low photon energies was used to observe the surface states present in the material. We have confirmed that the P-terminated surface shows spoon-like and bow tie-like shaped surface states whereas these shapes are not present in the Nb-terminated surfaces. In fact, the Fermi surface is so complex that one can observe different features of the Fermi surfaces along the $\overline{\Gamma} - \overline{X}$ direction with respect to the $\overline{\Gamma} - \overline{Y}$ direction. We also confirmed the band dispersion along $\overline{\Gamma} - \overline{X}$, as shown in Fig. 1 for both terminations, in an agreement with Ref. 22. This shows the distinct nature of the electronic states of two different cleaving planes. Both, P and Nb, terminations cannot coexist on the same cleaved surface because NbP has a single domain nature. The constant energy contour of the Nb-terminated surface is more complex than that of the P-terminated surface due to more Fermi surfaces around the projected Weyl points. As depicted in Fig. 2 (a) and 3 (a), bow tie and spoon-shaped features are mainly along $\overline{\Gamma} - \overline{X}$ and $\overline{\Gamma} - \overline{Y}$ directions but the 2D Fermi surface in the Nb-terminated plane is spread all over the surface Brillouin zone and different behavior along high symmetry directions is observed (see Fig. 5 (a)). The second derivative of the constant energy contour of Nb-terminated NbP consists of trivial and non-trivial Fermi pockets in which $S_5$ is a trivial circular hole pocket as well as $S_6$ is the trivial electron pocket in the middle region near $\overline{\Gamma}$ point. $S_7$ is a trivial hole pocket that is connected to the non-trivial pocket $S_8$ shown in Fig. 4 (a). It also shows two non-trivial Fermi pockets along $\overline{\Gamma} - \overline{X}$ direction out of which the Fermi arc $S_8$ is started and ended at a pair of W2 Weyl points. It is easy to detect this longer Fermi arc because it connects two different Brillouin zones ($\pm k_x$). $S_9$ is a non-trivial surface state along $\overline{\Gamma} - \overline{X}$ and connects to W1 Weyl points. Along $\overline{\Gamma} - \overline{Y}$ direction, small Fermi arcs connected to the Weyl points (W1 and W2) are in the same Brillouin zone [22].

### B. Modifications at Pb/NbP interface

We have in-situ evaporated approximately 1 ML of Pb on the P-terminated cleaned (0 0 1) surface of NbP. To verify the deposition of Pb, core level spectra have been taken after surface decoration (Fig. 2 (e).); Pb 5d related peaks confirm the deposition of Pb on the surface. Bow tie-shaped trivial surface states appear on the P-terminated surface due to the presence of dangling bonds after in-situ cleaving, whereas spoon-like features are topological SFAs. One monolayer of deposited Pb caused a dramatic change in the constant energy contours by affecting trivial as well as non-trivial surface states. Topologically trivial bow-tie shaped features were shrunken along $\overline{\Gamma} - \overline{X}$ and $\overline{\Gamma} - \overline{Y}$ directions due to dangling bond saturation, as demonstrated in Fig. 2 (b, d).

SFAs are the surface projections of bulk protected WPs which cannot be changed by deposition of foreign elements on the surface. However, one monolayer of Pb influenced the SFAs in such a way that they teleported from one pair of WPs to another one connecting two adjacent surface Brillouin zones and changed into a lemniscate shape (∞ shape) as depicted in Fig. 2 (f) [15]. The mirror plane for pure pristine P-terminated NbP is modified from $k_x$, $k_y = 0$ to $k_x$, $k_y = \pm \frac{\pi}{a}$ (see Fig. 2 (f)). This phenomenon is also known as the topological Lifshitz transition. The 5d electrons of deposited Pb have widely extended atomic orbitals hence it leads to strong hybridization with the P-terminated surface. In Fig. S1 [25], we plotted the constant energy contours for changing binding energy to show the evolution of the constant energy contours. This set of data proves that in the pristine surface, only spoon-like and bow-tie shaped features are present whereas after the surface decoration with 1 ML of Pb, lemniscate shape (∞ shapes) is present naturally along $\overline{\Gamma} - \overline{X}$ and $\overline{\Gamma} - \overline{Y}$ directions.

Furthermore, we observed modifications in the band dispersion and SFAs. Beyond a rigid shift just due to the chemical doping, we observed also a change in the connection of the WPs. To clarify this statement, we compared band dispersions along $\overline{M} - \overline{X} - \overline{M}$ and $\overline{M} - \overline{Y} - \overline{M}$. We observed a trident shape feature along these directions which is related to the change in the Fermi arcs shapes and connections (see Fig. 3). The band dispersion across the high symmetry path obtained for the pristine P-terminated NbP and after Pb deposition is consistent with the theoretical predictions [23], see Fig. 2 (c, d).

In the next trial, we freshly cleaved another crystal of NbP perpendicularly to [0 0 1] direction and observed the ARPES spectra for Nb-terminated NbP and decorated it with approx. 1.9 ML of Pb. We confirmed the presence of a Pb layer by using core-level spectra, as shown in Fig. 4 (b), indicating two significant peaks of Pb 5d orbitals. The deposition of 1.9 ML of Pb on the Nb-terminated surface of NbP altered only trivial surface states while SFAs remained unchanged. Four new additional trivial pockets appeared with pre-existed trivial hole pocket ($S_6$) along $\overline{\Gamma} - \overline{M}$ directions (see Fig. 4 (c)) which is due to the Fermi surface of Pb deposited on Nb-terminated NbP [24]. We compared band dispersions along high symmetry paths, as shown in Fig. 5. Additionally, we observed the Fermi surfaces connected to $S_8$ along the vertical direction. After careful analysis of constant energy contour changing with binding energy (see Fig. S3, [25]), it is shown that non-trivial SFAs remained the same with no change. Hence, the transition is an ordinary Lifshitz transition, which is just a change in the Fermi surface without changing the connections between the Weyl points.

### C. Modifications at Nb/NbP interface

We have deposited 0.8 ML of Nb on the P-terminated (0 0 1) plane. Nb decorated P terminated surface along the $\overline{M} - \overline{X} - \overline{M}$ and the $\overline{M} - \overline{Y} - \overline{M}$ k-path, shown in Fig 6, looks very much similar to a constant energy contour in Fig 5 (c). from Ref. [15] which supports the fact that 0.8 ML Nb deposited on P-terminated NbP is at the transition point of topological Lifshitz transition but not yet fully transformed. At this transition, the trivial surface states are contracted but not fully converted into lemniscate shape (∞-like shape). Cuts along $\overline{X} - \overline{\Gamma} - \overline{X}$ and $\overline{Y} - \overline{\Gamma} - \overline{Y}$, demonstrated in Fig. 6 (g, h, i, j), show a linear dispersion proving the existence of Weyl points before and after the Nb deposition on pristine P-terminated NbP. This amount of deposition modifies the electronic structure of NbP where trivial bow-tie shaped features vanished after the surface decoration due to dangling bond saturation (see Fig. 7). The constant energy contour with changing binding energy is shown in supplementary Fig. S2 [25] to illustrate the change in the Fermi surface after the deposition of guest Nb atoms.

We also tried to decorate Nb terminated NbP with Nb. Such a procedure resulted in the appearance of blurred, features due to the disorder introduced in the system.

The summary of the experiments done in this article is shown in the form of 3D intensity plots where constant energy contours and corresponding band dispersions are depicted together (see Fig. 8.)

## IV. CONCLUSIONS

The studied NbP single crystal was cleaved in-situ to obtain (0 0 1) surfaces with two possible terminations (P and Nb, respectively). ARPES studies showed distinct natures of the electronic states of two different cleaving planes. The Nb-terminated surface exhibits trivial and non-trivial surface states features that are different along the $\overline{\Gamma} - \overline{X}$ and $\overline{\Gamma} - \overline{Y}$ directions. We can summarize the facts as follow:

1. We were able to observe the strong modifications of the trivial and non-trivial surface states due to topological Lifshitz transition in Pb deposited P-terminated NbP with the modifications in the SFAs, while the WPs remained robust against the change of the surface environment. The robustness of the WPs was already demonstrated in the literature for a cover of light elements. In this paper, we go beyond demonstrating that the WPs are robust even against a cover with heavy elements, which produce a stronger perturbation of the electronic structure since they are characterized by larger spin-orbit coupling and larger strength of the electronic hybridization. The WPs are so robust since they rise from the topology of the bulk, therefore they are barely affected by the cover on the surface.

2. For P-terminated NbP, we observed that the deposition of about 0.8 ML of Nb was not enough to fully transform the system according to TLT, however, the system reached the Lifshitz transition point.

3. For Nb-terminated NbP covered with 1.9 ML of Pb, ordinary Lifshitz transition occurred with unchanged topological SFAs and additional trivial pockets occurring due to individual features of Pb and Nb-terminated NbP.

4. Deposition of Nb on the Nb-terminated NbP surface led to a formation of a disordered system with blurred features poorly resolved in the APRES results.

## ACKNOWLEDGEMENTS

The work is partially supported by the Foundation for Polish Science through the International Research Agendas program co-financed by the European Union within the Smart Growth Operational Programme. This publication was developed under the provision of the Polish Ministry of Education and Science project: "Support for research and development with the use of research infrastructure of the National Synchrotron Radiation Centre SOLARIS" under contract nr 1/SOL/2021/2. We acknowledge SOLARIS Centre for the access to the Beamline UARPES, where the measurements were performed.

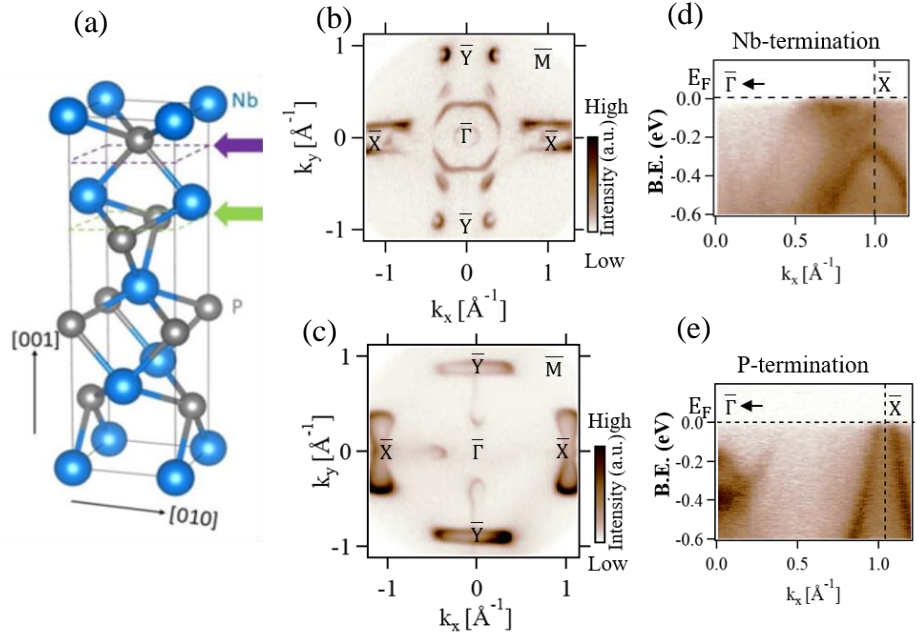

Fig. 1. (a) Crystal structure of NbP with two possible cleavage planes shown by two different color arrows, ARPES images of (b) Nb-terminated (c) P-terminated NbP surfaces. (d, e) band dispersion along $\overline{\Gamma}$-$\overline{X}$ direction for the Nb- and P- terminations, respectively.

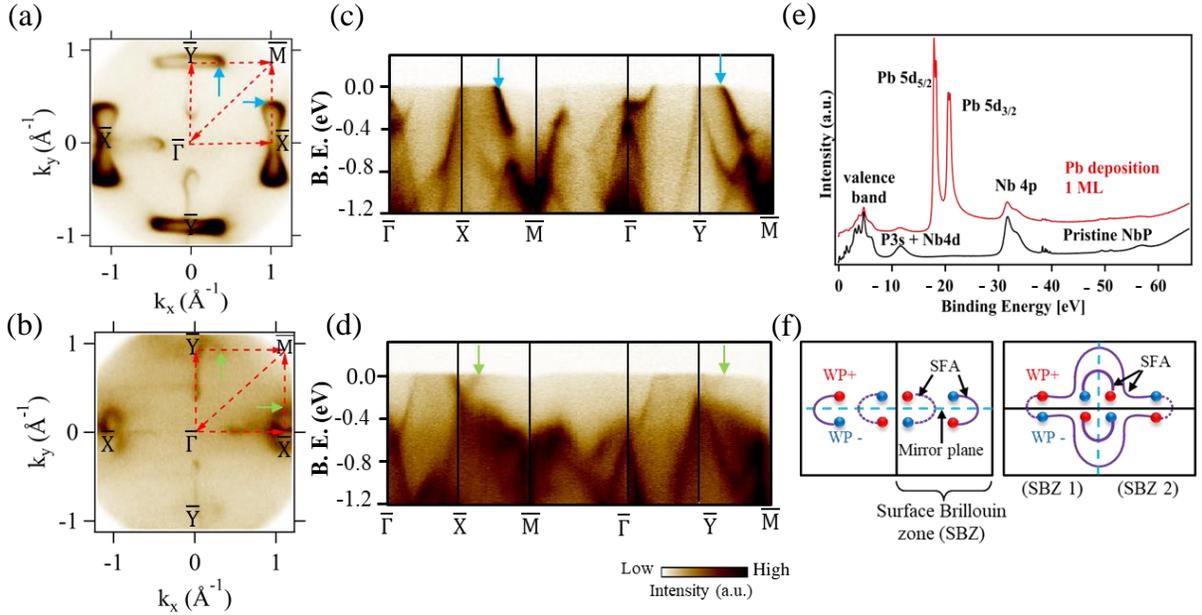

Fig. 2. Pristine P-terminated NbP surface and its modification after Pb evaporation: (a) constant energy contour of pristine NbP with high symmetry path shown (red color) and trivial surface states shown by blue arrows, (b) surface modifications after Pb deposition with high symmetry path in which the nontrivial surface states are indicated by green arrows, (c, d) corresponding band structure along the same high symmetry path with topological surface states (TOPOSS), (e) core level spectra of pristine NbP and Pb deposited on P-terminated NbP, (f) the figure is based on the figure taken from Ref. [15] and illustrates the surface Brillouin zone with SFAs and their connections to WPs *I suggest to use also acronym WP- (as is used in Nature paper, in analogy to WP+ and for completeness explain the meaning)* before and after Pb deposition. For pristine P-terminated NbP, the mirror plane is shown by horizontal dashed blue line shown by an arrow which changed as vertical blue dashed line after 1 ML Pb deposition.

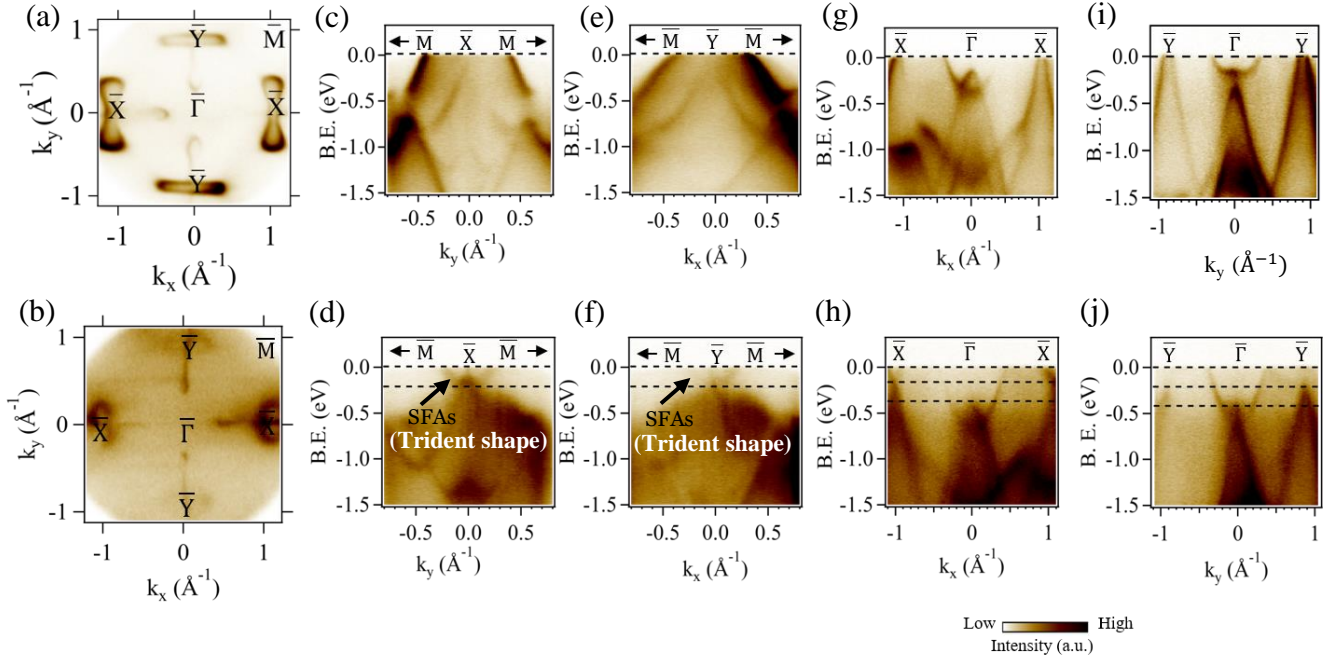

Fig. 3. Comparison between pristine P-terminated NbP and surface modifications after 1 ML of Pb deposition on it, constant energy contour of: (a, b) Pure P-terminated NbP and modified P-terminated NbP, respectively, comparison of band dispersions along (c, d) $\overline{M} - \overline{X} - \overline{M}$ (e, f) $\overline{M} - \overline{Y} - \overline{M}$, (g, h) $\overline{X} - \overline{\Gamma} - \overline{X}$, (i, j) $\overline{Y} - \overline{\Gamma} - \overline{Y}$.

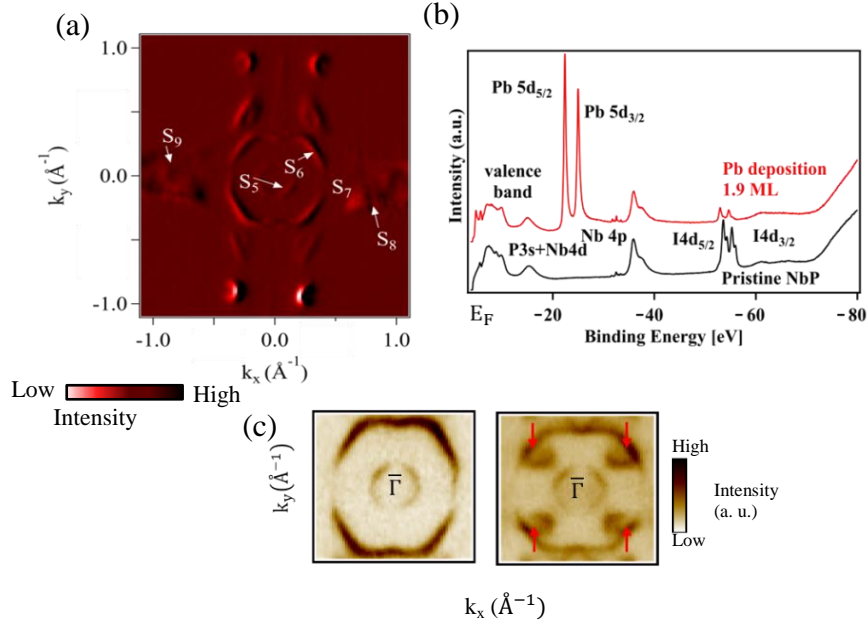

Fig. 4. Features of the Nb-terminated NbP: (a) second derivative of 2D Fermi surface with denoted the Fermi arcs to show clear fingerprints of trivial and non-trivial Fermi surfaces (in which $S_5 - S_7$ are trivial surface states and $S_8$, $S_9$ are non trivial surface states), (b) core level spectra of pristine NbP and the same surface decorated with 1.9 ML of Pb with two significant peaks showing 5d orbitals (in this particular crystal, Iodine peaks: because Iodine was used as a transport agent during the crystal growth process), (c) comparison of ARPES spectra near $\overline{\Gamma}$ points before and after Pb deposition showing new additional four pockets introduced to energy contour marked by red arrows.

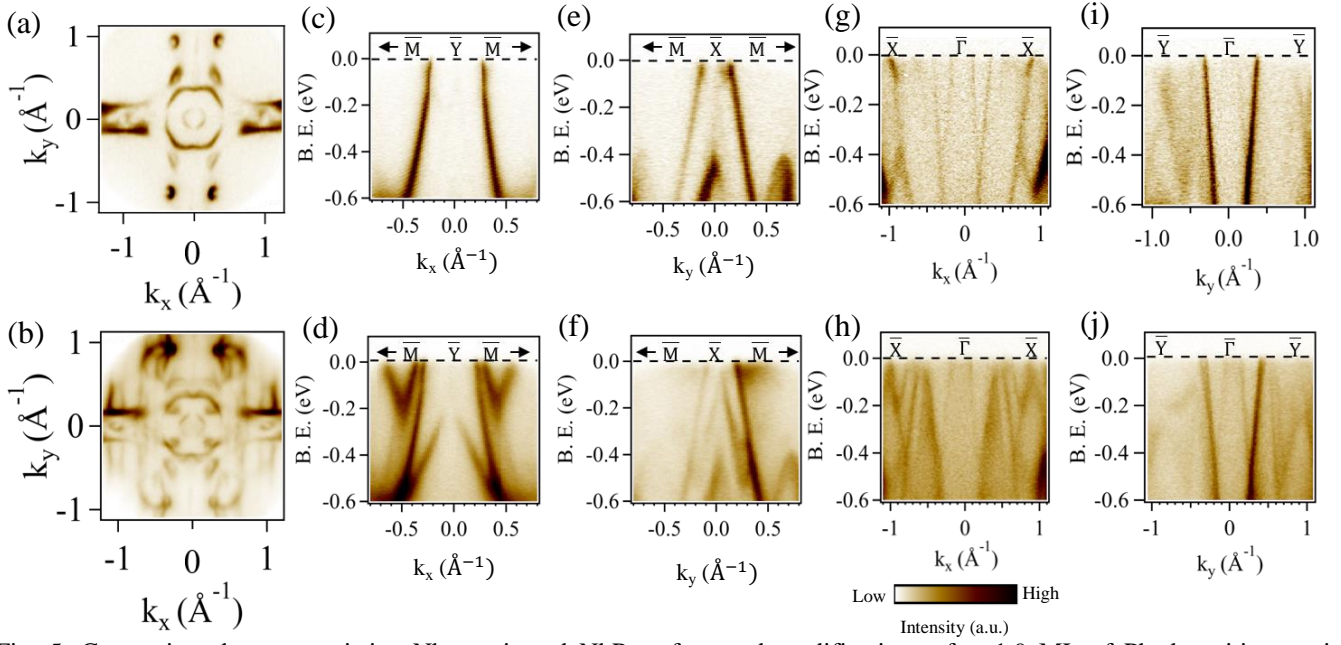

Fig. 5. Comparison between pristine Nb-terminated NbP surface and modifications after 1.9 ML of Pb deposition on it, constant energy contour of: (a, b) pure Nb-terminated NbP and modified Nb-terminated NbP, respectively, comparison of band dispersions along: (c, d) $\overline{M}-\overline{Y}-\overline{M}$, (e, f) $\overline{M}-\overline{X}-\overline{M}$, (g, h) $\overline{X}-\overline{\Gamma}-\overline{X}$, (i, j) $\overline{Y}-\overline{\Gamma}-\overline{Y}$.

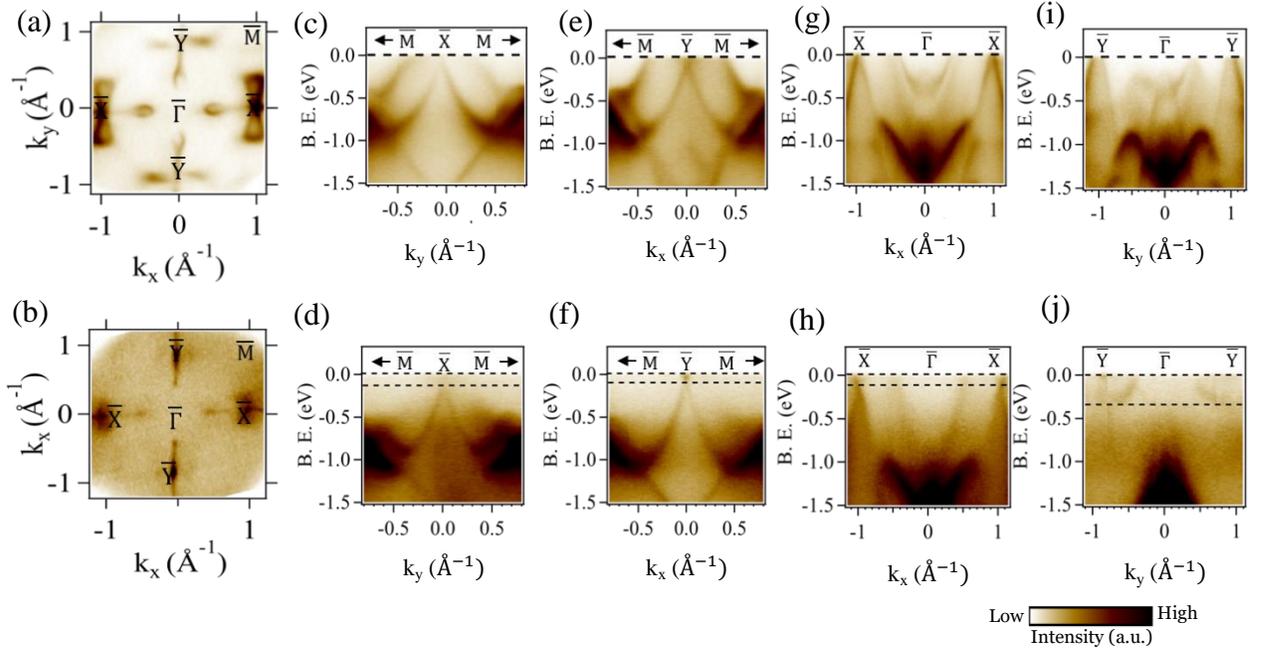

Fig. 6. Comparison between pristine P-terminated NbP and surface modifications after 0.8 ML of Nb deposition on it, constant energy contour of: (a, b) pure P-terminated NbP and modified P-terminated NbP respectively, comparison of band dispersions along: (c, d) $\overline{M}-\overline{X}-\overline{M}$, (e, f) $\overline{M}-\overline{Y}-\overline{M}$, (g, h) $\overline{X}-\overline{\Gamma}-\overline{X}$, (i, j) $\overline{Y}-\overline{\Gamma}-\overline{Y}$.

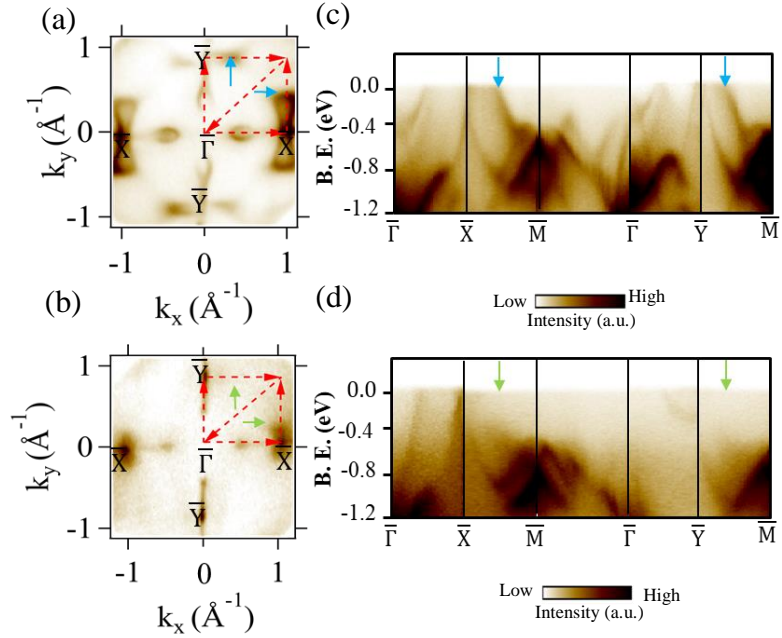

Fig. 7. Pristine P-terminated NbP in comparison with Nb decorated P-terminated NbP: (a, b) 2D Fermi intensity plot of pure P-terminated NbP surface and 0.8 ML deposition of Nb on it. High symmetry paths are indicated by red dotted arrows. (c, d) the band structure before and after 0.8 ML of Nb deposition, obtained along $\overline{\Gamma} - \overline{X} - \overline{M} - \overline{\Gamma} - \overline{Y} - \overline{M}$ path for pure P-terminated NbP and Nb deposited NbP. Blue arrows in (a) and (c) indicating trivial bowtie shaped surface states before surface decoration whereas green arrows in (b) and (d) shows the same states vanished after Nb deposition.

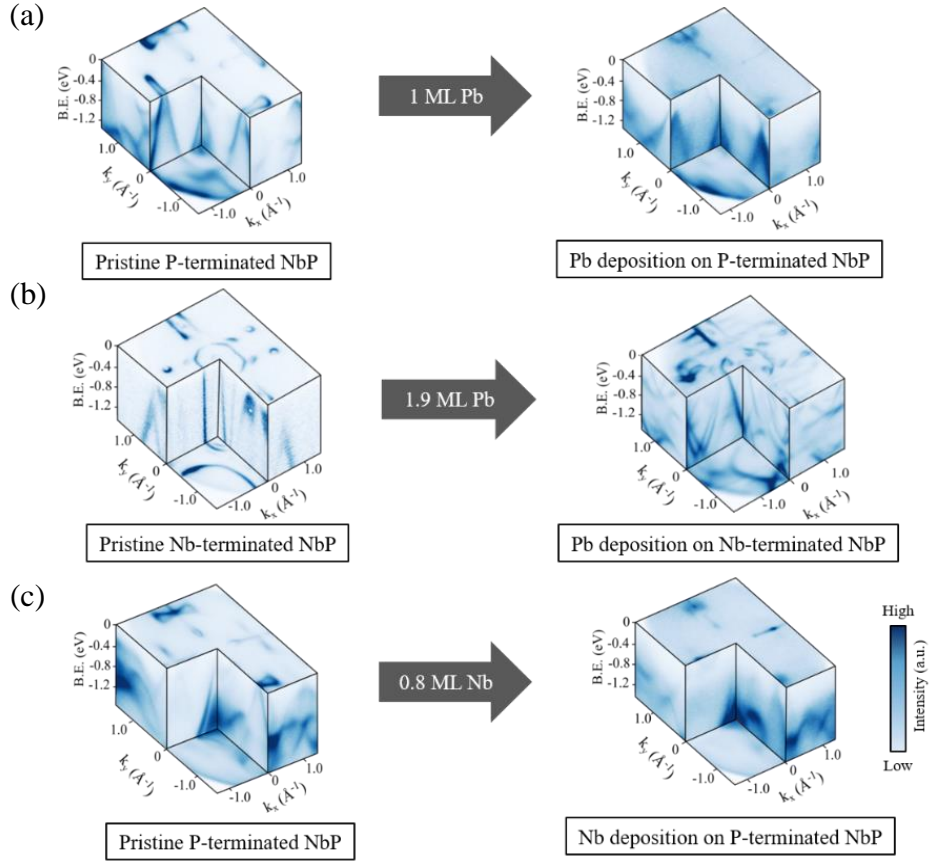

Fig. 8. The comparisons between 3D intensity plots of: (a) pristine P-terminated NbP as well as 1 ML of Pb deposited on P-terminated NbP, (b) pristine Nb-terminated NbP and modifications obtained after 1.9 ML Pb deposition, (c) pristine P-terminated NbP and modification obtained after 0.8 ML Nb deposited on P-terminated NbP

8